# Electronic and structural distortions in graphene induced by carbon vacancies and boron doping

Ricardo Faccio[a,c(*)], Luciana Fernández-Werner[a], Helena Pardo[a,c], Cecilia Goyenola[a], Oscar N. Ventura[b] & Álvaro W. Mombrú[a,c].

[a]Crystallography, Solid State and Materials Laboratory (Cryssmat-Lab), DETEMA, Facultad de Química, Universidad de la República, Gral. Flores 2124, P.O. Box 1157, Montevideo, URUGUAY & Centro Nano*Mat*, Polo Tecnológico de Pando, Facultad de Química, Universidad de la República, Cno. Aparicio Saravia s/n, 91000, Pando, Canelones, URUGUAY

[b]CCBG, DETEMA, Facultad de Química, Universidad de la República, URUGUAY

[c]Centro Interdisciplinario en Nanotecnología y Química y Física de Materiales, Espacio Interdisciplinario, Universidad de la República

(*) corresponding author: rfaccio@fq.edu.uy

**Abstract**

We present an *ab initio* study on the structural and electronic distortions of modified graphene by creation of vacancies, inclusion of boron atoms, and the coexistence of both, by means of thermodynamics and band structure calculations. In the case of coexistence of boron atoms and vacancy, the modified graphene presents spin polarization only when B atoms locate far from vacancy. Thus, when a boron atom fills single- and di-vacancies, it suppresses the spin polarization of the charge density. In particular when B atoms fill a di-vacancy a new type of rearrangement occurs, where a stable $BC_4$ unit is formed inducing important out of plane distortions to graphene. All these findings suggest that new chemical modifications to graphene and new type of vacancies can be used for interesting applications such as sensor and chemical labeling.

## I. INTRODUCTION

Carbon based materials and its related nanostructures have been the target of many studies in different fields such as chemistry, physics and biology, because of their interesting fundamental properties, and their potential use in technological applications. One important topic corresponds to the electronic structure of carbon materials in general, because of the high coherence and high charge conductivity they present[1], among other relevant properties. Additionally, carbon based materials work as testing bench demonstrating that metal free sp-compounds present magnetic properties such as ferro- and ferrimagnetism[2].

Regarding magnetism, many works focused in the production of magnetic carbon from different routes: chemical vapor routes[3,4], ion-bombardment[5,6], nanofoam[7], ion implantation, etc[2]. First Principles calculations have been used many times for revealing the role of defects or vacancies for inducing magnetism in carbon. One of the first reports on vacancies in graphene is the work of Lehtinen *et al*[8]. After that, we extended the work to graphite considering the coupling of aligned vacancies along c-axis of graphite[9] that differs for graphene in the magnetic moment per carbon atom, introducing structural modifications. Another example corresponds to Zhang *et al*[10], who have found that total magnetization in diamond and graphitic structures decreases when the vacancy density increases, while the presence of nitrogen atoms near a vacancy enhance the magnetic signal. The importance of this lies in the possibility of tuning magnetization by defect generation and doping. In 2009 Červenka *et al*[11] reported direct experimental evidence for ferromagnetic order locally at defect structures in highly oriented pyrolytic graphite by Magnetic Force Microscopy and bulk magnetization measurements at room temperature, and using our 3-D model[9] for describing the coupling between different vacancy-to-vacancy distances at edges. More recently Ugeda *et al*[12] have experimentally observed the presence of a sharp electronic resonance at the Fermi energy around each single graphite vacancy by scanning tunneling microscopy experiments, confirming the predictions from DFT and tight-binding around the presence of localized states, as a source of magnetism in carbon nanostructures.

Regarding chemical modification, doping in graphene offers the possibility to enlarge the wide range of applicability of this material. For this reason, big efforts are being made in order to prepare, characterize and understand doped graphene. Dai *et al*[13] have performed a theoretical study of the gas adsorption on graphene doped with different elements, suggesting that B- and S-doped graphene could be a good sensor for polluting gases such as NO and $NO_2$. In another work, Wang *et al*[14] have fabricated an n-type graphene field-effect transistor which operates at room temperature by edge functionalized graphene nanoribbon (GNR) with nitrogen species using high power electrical annealing in $NH_3$. Simultaneously, they have theoretically studied the edge functionalization of GNRs by oxygen and nitrogen-containing species, confirming the possibility of p- and n-doping GNRs (up to a width of 40nm) respectively. In addition, B-and N-doped graphene have been prepared and characterized by Panchakarla *et al*[15]. B-doped graphene was synthesized by arc discharge of graphite electrodes in presence of $H_2+B_2H_6$ and also by arc discharge of boron-stuffed graphite electrodes while N-doped graphene was obtained by arc discharge in the presence of $H_2$+pyridine or $H_2$+ammonia, as well as performing the transformation of nanodiamond in the presence of pyridine. They concluded that B- and N-doped graphene can be synthesized to exhibit p- and n-type semiconducting electronic properties that can be systematically tuned with the concentration of B and N. These works confirm that doped-graphene is actually being experimentally obtained by diverse techniques and for different purposes, despite the fact of the general difficulties of doping in nanostructures. Actually, this is the subject of many theoretical investigations in order to find better-controlled methods. Pontes *et al*[16] proposed a barrier-free substitutional doping of graphene sheets with boron atoms, after performing their *ab initio* calculations. They placed graphene between two media, one of them B-rich and the other one N-rich. In this way a C atom is pushed out from the layer and a B atom is incorporated in its place. In a second step the C-N radical bonded to the B atom is removed inserting the system in an H-rich medium. They also showed that the B atom could be incorporated when N atom is substituted by P, O, CH, and NH groups. The authors deduced that this mechanism could explain the B doping of carbon nanotubes performed by X. M. Liu and co-workers obtained by exposure of single wall carbon nanotubes to $B_2O_3$ and $NH_3$ [17].

Thus, doping graphene simultaneously with the presence of vacancies constitutes a large step towards understanding of the electronic structure of metal free systems, with properties that can be tuned by chemical and structural modifications. For this reason we propose to work altogether with boron doping and vacancies in graphene, discussing thermodynamics, structural and electronic consequences of the mentioned chemical/structural modifications. To our knowledge this could be the first work discussing coexistence of boron atoms with vacancies in graphene.

The paper is organized as follows. Methods are presented in section II. Results and discussion is divided in III-i, III-ii and III-iii, corresponding to: vacancies in graphene, b-doping of graphene and coexistence of single carbon atom vacancies with boron doping respectively. Finally, conclusions are presented in section IV.

## II. METHODS

The electronic structure simulation were based on the First Principles – Density Functional Theory[18,19] which we have successfully used to study bulk graphene, thioepoxidated SWCNT, sulfur doped graphene and double wall carbon nanotubes and graphene nanoribbons[9,20,21,22,23]. The simulations were performed using SIESTA code [24,25,26] which adopts a linear combination of numerical localized atomic-orbital basis sets for the description of valence electrons and norm-conserving non-local pseudopotentials for the atomic core. The pseudopotentials were constructed using the Trouiller and Martins scheme[27] which describes the interaction between the valence electrons and the atomic core. We selected a split-valence double-ζ basis set with polarization orbitals for all the atoms. The extension of the orbitals is determined by a cutoff radii of 4.994 a.u. and 6.254 a.u. for *s* and *p* channels respectively, as obtained from an energy shift of 50 meV due to the localization, with a split norm of 0.15. The total energy was calculated within the Perdew–Burke–Ernzerhof (PBE) form of the generalized gradient approximation GGA xc-potential[28]. The real-space grid used to represent the charge density and wavefunction was the equivalent of that obtained from a plane-wave cutoff of 250 Ry. For modeling graphene we took the coordinates of one carbon layer of graphite, and then we expanded the *c*-axis up to 10 Å, in order to avoid the interaction between images in adjacent cells. For describing pure graphene (gxx), vacancies in graphene (gxx+v), b-doped graphene (gxx+b) and b-doped graphene with vacancies (gxx+bv) we construct supercells defined in terms of pristine graphene. The atomic positions were fully relaxed in all the cases using a conjugate-gradient algorithm until all forces were smaller than 0.02 eV/Å. In addition the unit cells were relaxed until the components of the Voigt's tensor were smaller than 0.2 kBar. For the k-points sampling of the full Brillouin Zone we selected Monkhorst Pack grids[29] ranging from 40x40x1 to 100x100x1 for describing the 6x6 to 2x2 graphene supercells respectively. All these parameters allow the convergence of the total energy and forces. The detailed information of different cases is explained along the text.

## III. RESULTS AND DISCUSSION
### III. i. Single atom vacancies on graphene

The first step towards understanding the coexistence between vacancies and boron is to understand the existence of vacancies itself. We performed several calculations for vacancy concentrations ranging from 1.4% to 33.3%, by means of supercells, defined in terms of pristine graphene. The nomenclature for the single atom vacancy in graphene corresponds to: gxx+v; where "x" stands for the number of supercells along the *a* and *b* axis, while "v" indicates the presence of a single vacancy. The geometrical reconstruction was performed minimizing both forces and tensor stress components, as was mentioned in section II.

The total magnetic moment per supercell changes respect to the defect concentration, with a smooth oscillation as presented in Table 1. The highest net magnetic moment per supercell was obtained for g55+v, while the lowest one corresponds to the g22+v case. However, the opposite trend was obtained when normalizing with respect to the quantity of carbon atoms constituting the super cells. The observed fluctuations of the net magnetic moment indicate that interaction between vacancies images are still present, and work in some fashion where localization versus delocalization of the spin density plays an important role. In order to shed some light on this we can analyze the g55+v case. The decomposition of the density of states indicates that the dangling $sp^2$ orbital is located about c.a. 1 eV below the Fermi level, contributing with a magnetic moment of 1.0 $\mu_B$. The remaining magnetic moment, m=0.34 $\mu_B$, corresponds to the unbalance between spin polarized π-orbitals. Since these levels are quite proximate to the Fermi level, they become more susceptible to structural changes, and because of their delocalization they extend more than flat $sp^2$-states (or σ-), spreading through the whole structure. All these features can be seen in Figure 1, where the spin polarized charge density is plotted for the corresponding charge decomposition, σ- and π-bands.

The oscillation in the net magnetic moment for the different supercells can be understood considering the magnetic moment carried by the $sp^2$ orbital as a constant background of c.a. 1 $\mu_B$, with an additional fractional moment originated in the unbalance of the partial occupied π-bands. This feature is strongly dependent on the defect concentration. That's where oscillation comes, because the localization is highly affected for the band depopulation introduced by the presence of vacancies that behave according to the Stoner's band magnetism model[30]. This trend has been seen in the work of Yazyev *et al*[31] obtaining similar results to ours. Additionally, the spin arrangement corresponds to a ferrimagnetic ordering, with opposite spins for atoms in different sub-lattice position.

Additionally, the formation energy of a vacancy for different concentration was determined. For this purpose we use the following expression:

$$\text{Pure-Graphene}_{gxx} \rightarrow \text{Vacancy-Graphene}_{gxx+v} + C_{bulk}$$

$$E_{Form}(1) = E_{vacancy} = E_{gxx+v} - N\mu[C]$$

where $E_{gxx+v}$ and $\mu[C]$ corresponds to total energy of a single vacancy in graphene and the chemical potential for carbon respectively, and N indicates the total number of carbon atoms in the defective case. The chemical potential for graphene was determined as the energy per atom in the case of pristine material $\mu[C]=-156.223$ eV, which was checked from g11 up to g66 obtaining differences in the energy c.a. 1E-3 eV. The formation energies obtained from these values, which are listed in Table 1, have shown a small oscillation around 7.40(15) eV, which agrees quite well with other reports[32,33,34]. In order to evaluate the accuracy of the basis set we gave more flexibility using additional "floating basis", by placing empty 2s and 2p carbon orbital centered in the removed atom position. In this case the augmentation of the basis set introduces negligible changes in the magnetic moment of the system, with changes in less than 1%. On the other hand, the total energy of the system is decreased in about 0.05 eV. This value can be taken as a measurement of the BSSE introduced in the method.

The case of g22+v is an exception, since the formation energy is only slightly lower than the value obtained for the remaining cases, but this is consistent with the exception of what g22+v introduces because of its high density of defects. This feature produces a Stoner band splitting not only for π-bands, but for σ-bands too, as can be seen in the corresponding density of states presented in Figure 2, establishing an important difference in respect to the other used supercell.

It is important to note that non-magnetic solution is always higher in energy than the magnetic one, for about 0.4 eV. This feature is common for the whole serie, and is accompanied by a structural distortion, in which the under-coordinated carbon atom, located at position 3 (Figure 2), goes into a Jahn-Teller distortion[35], moving out from the graphene plane in 0.3 Å. Starting from this last structure, a further spin polarized calculation breaks the out-of-plane distortion, relaxing to the original flat-graphene structure with a net magnetic moment, with identical results to the ones listed in Table 1. All this is a confirmation of the structural and electronic stability of the proposed magnetic ground state solution.

### III. ii. Boron doped graphene

The next step consists in filling the carbon atom vacancies with boron atoms, starting by placing B atoms in different positions and allowing the structure to relax. While the in-plane position seems to be the most stable structure for this system, we checked different out-of-plane starting positions for B, but in all the cases the structures relax to the same in-plane position. In this case the B-doped graphene retains most of the hexagonal symmetry of graphene, with slight changes in the lattice parameters in reference to graphene. For high boron concentration the lattice expands, and then it contracts for low concentration, as can be seen in Table 2. This trend can be easily explained considering that the B-C distance remains almost constant (c.a. 1.51 Å), and its value is higher than the C-C distance, so that a higher concentration of B will contribute to the lattice expansion of the original graphene supercell. All these features indicate, from a strict structural point of view, that similarities between B and C atoms are enough to retain most of the graphene symmetry. In this case the point group at the boron site corresponds to $3/m \equiv \bar{6}$, according to the space group $P\bar{6}m2$. While the point group at B is adequate for describing the whole range of the studied doping, in the case of high concentration of defects the unit cells show smalls changes, in particular at the γ angle with deviations of 0.02° that can be considered as a negligible value for breaking the hexagonal symmetry of the space group. All these features correspond only to the structural perspective of the system that cannot be further extended to its electronic structure, as we will discuss later.

The thermodynamics of the doping indicates that for a solid state reaction between graphene and solind B we can sketch the reactions and the corresponding formation energy as:

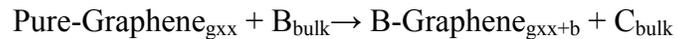

$$\text{Pure-Graphene}_{gxx} + B_{bulk} \rightarrow \text{B-Graphene}_{gxx+b} + C_{bulk}$$

$$E_{Form}(2) = E_{gxx+b} - N\mu[C] - \mu[B]$$

where $E_{gxx+b}$ corresponds to the total energy of the b-doped graphene supercell, the chemical potential for B was determined as the energy per atom from its rhombohedral structure[36] as $\mu[B] = -77.06$ eV. On this basis the formation energy for the B doping of perfect graphene is $E_{Form}(2)=1.0(1)$ eV, which is a non spontaneous process but considerably lower in energy than the vacancy creation process for graphene. It is important to note that the formation energies has a negative slope regarding the size of supercell, indicating that the doping by highly diluted B will be the most favorable process, as can be seen in Table 2.

Additionally, we can also consider another situation where boron could be incorporated, when starting from a defective graphene:

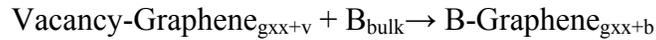

$$E_{Form}(3)=E_{gxx+b}-E_{gxx+v}-\mu[B] = E_{gxx+b} - \{E_{Form}(1)+N\mu[C]\} - \mu[B]=E_{Form}(2)-E_{Form}(1)$$

Now the formation energy turns more favorable, since defective graphene contains high energetic dangling bonds that react immediately with boron. From this point of view the chemical doping of defective graphene can be seen as a highly favorable reaction, with a mean formation energy of $E_{Form}(3)= -6.4(2)$ eV, and the fact that diffusion of boron into the graphene plane is a barrier-free process, as was mentioned earlier and sketched in Figure 4, confirms that this process is kinetically favorable too.

The electronic structure of graphene changes due to the boron incorporation, because p-doping modifies graphene into a metal. A Mulliken's population analysis[37] presented in Table 2 indicates that boron atoms get electron from its carbon neighbors, generating charge depletion in their neighborhoods. This feature is sketched for g66+b in Figure 3, where the first coordination shell shows carbon atoms with a net charge of c.a. +0.17 $e^-$ each, and then diminished for further carbon atoms shells. Boron atoms get at most 0.95 electrons from graphene in the most B diluted case, and 0.80 electrons for the most B concentrated case. In principle 50% of the additional electron located at boron comes from the first carbon atom shell, while the remaining 50% comes from the rest of the atoms which carried net charges per atom in the range of +0.03 e$^-$ to +0.004 e$^-$. Since g22+b is the most concentrated supercell it has just a few of these slightly charged carbon atoms, the net charge transfer is limited to the closest carbon atoms to the boron center obtaining just 0.90 electrons. The opposite occurs for g66+b, where there are many more atoms, allowing boron to obtain as much as 0.95 electrons from its neighbors. For this reason, the limit in the net charge of boron comes from the number of independent atoms in the supercell.

As was mentioned before, the electronic properties of graphene become altered due to the incorporation of B, inducing a p-hole doping turning graphene in metal. According to Mulliken the induced net charge has always a $p_z$ character. All this indicates important changes at π-bands, in particular a band depopulation that can be tuned playing with the concentration of boron in the structure. With the aim to confirm this we performed a charged supercell calculation including an additional electron per boron atom in the supercell. This electron compensates the introduced hole obtaining an iso-electrical-graphene. After the incorporation of the additional electron we recover an insulator behavior for g22+b+e, g44+b+e and g55+b+e where a gap of c.a. 0.7 eV is open at Dirac point, while for g33+b+e and g66+b+e, we obtain zero gap behavior. All this can be seen in Figure 5. The reason for these differences arises in the band folding, which is different from g33/g66 in comparison with the other supercells, since the Dirac point K, originally located at $(2\pi/a_{pristine})(1/3,1/3,0)$ for pristine graphtite, now moves into the Γ point of the new Brillouin zone $(2\pi/a_{supercell})(0,0,0)$ for g33/g66, because we have a 3n factor in the new direct cell parameters. Additionally the highest occupied π-bands become twofold degenerated for g33 and g66 pristine graphene. When B⁻ substituted one carbon atom the twofold degeneracy is broken, one of them with a π-π* contact at Dirac point and the other one opens up a pseudo gap as can be seen in for g66+b. The reason for that splitting can be easily seen by inspection of both wavefunction plotted at Dirac point presented in Figure 5. Here we can see how the wavefunction becomes distorted for lobules located near the B position, while for the nodeless wavefunction at B there is no distortion and the band presents similar dispersion to the pristine wavefunction.

All the mentioned aspects that were presented are a direct consequence of the differences in the crystalline potential introduced by boron, which breaks the electronic symmetry of pure graphene, but this anisotropy is not sufficiently for induce localized states at the Fermi level. For this reason the density of states shows always a smooth slope, justifying a non-spin polarized ground state. As a consequence, for the whole range of B doping, filled vacancies are not magnetic at all.

### III. iii. Coexistence of single atom vacancies and boron doping

So far we have discussed the effect of vacancies and boron doping independently. Now we are going to enter into details considering the coexistence of both, which is a complex task, since we have to consider several configurations. For this reason we just limited all the possibilities to three different vacancies to boron distances[38], which are show in Fig. 6 for 5x5 supercell. In the first case we considered the carbon substitution by boron atoms in graphene. We doped with boron atoms at the origin of the supercell, while the vacancy is located at the center of the supercell, with the longest boron to vacancy distance (gxx+bfv: boron far from vacancy). Then we simulated the boron doping adjacent to the vacancy, substituting one of the carbon atoms at the saturated pentagonal bond (gxx+bav_a: boron adjacent to vacancy, with asymmetric disposition), corresponding to position 1 and 2. Finally, we considered the other doping adjacent to the vacancy, in which the substitution is performed over the under-coordinated carbon atom (gxx+bav_s: boron adjacent to vacancy, with symmetric disposition). The reason for using the term symmetric and asymmetric is originated in the final structural reconstruction for one of the boron type of doping, in which boron atom spontaneously moves into the centre of the vacancy, establishing a fourfold coordination environment for this atom tetrahedral-like $BC_4$ unit. In this structure boron establishes an unusual coordination polyhedron for graphene, being this structural reconstruction capable of driving an out-of-plane distortion on the modified graphene.

Formation energy for this doping can be described according to the following reaction:

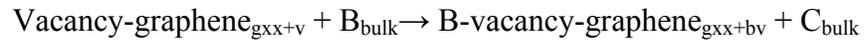

Vacancy-graphene$_{gxx+v}$ + B$_{bulk}$ → B-vacancy-graphene$_{gxx+bv}$ + C$_{bulk}$

$$E_{Form}(4) = E_{gxx+bv} + \mu[C] - E_{gxx+v} - \mu[B]$$

In this case we started from a defective graphene where boron incorporates into the lattice removing one carbon atom that goes into another place in graphene. The results are presented in Table 3, where energy values and magnetic moments are listed for all the cases. The most stable configuration corresponds to the fourfold coordinated B atom which saturates all the dangling carbon atom bonds, adopting a distorted-tetrahedral geometry. Total energy of supercells decreases when boron atoms occupy closer position to the vacancies. This result is quite obvious, since dangling bonds are highly energetic, and the saturation of all bonds clearly stabilizes the structure. For this reason formation energy indicates a favorable process for about c.a. 1.5 eV, but the process becomes unfavorable for about c.a. 6.4 eV when pure graphene is considered as a reactant. Table 3 presents the mentioned parameters, together with $E_{norm}$ consisting in the direct difference in total energy in comparison the lowest total energy, allowing a quick comparison for same supercells. According to this, bulk boron doping is unfavorable in reference to gxx+bav_s for about 2.6 eV, while doping in gxx+bav_a is unfavorable in 0.3(1) eV. While the last value seems to be low, it is sufficient to drive the formation of a distorted tetrahedral $BC_4$ unit.

Regarding the electronic structure, when boron fills vacancies, a smoothing of the density of states that avoids any spin-polarization of the charge density is revealed. In essence a carbon di-vacancy can be easily cured by a single atom B that goes at the middle of the vacancy, inducing an out-of-plane distortion, with a zero magnetic moment. Concerning stability for this particular conformation we performed two checks. One of them consists in relaxing the actual coordinates with another basis set, such as plane waves, using Quantum Espresso[39]. The structure relaxed to same atom positions, with cell parameters differences less than 0.5% and a non-spin polarized solution was obtained too[40]. Besides that, we determined the phonon density of states using SIESTA code, in order to check for positive frequencies. A 3x3 supercell of the g44+bav_s was used to determine the force constant for every single atom of the central g44+bav_s region, in order to get the force constant matrix, which is further submitted to a Fourier transform in order to get phonons eigenvectors and frequencies. After this procedure we obtained all positive frequencies that are presented in Figure 6 by its vibrational density of states, weighted for both carbon and boron atoms. All the tests, confirmed that gxx+bva_s constitutes a stable solution.

Since the B-C reconstruction in gxx+bav_s is particularly unusual for boron in this kind of structures, we performed further calculations using other atoms, similar in radius, such as: N, Li and Be obtaining different results than B. In all the cases, central atom moves to form a hexagon, leaving an under-coordinated carbon atom at position 3. Finally, we performed a calculation including a carbon atom in the central position of boron, the reconstruction goes similar to the one described above, but it follows with a further out of plane distortion of 0.28 Å for the under-coordinated carbon atom (position 3). The reason for that change is explained in terms of the lack of the spin polarization of this system. Since the original four-fold coordinated atom turns spontaneously non-polarized, the system kept in that spin configuration until force tolerance is reached. It is clear that single atom vacancies reconstruction in graphene presents quite different results for spin polarized electrons and non-spin polarized electrons as was mentioned in section III.i. If we take the last obtained coordinates and then we reinitialize the density matrix but with a spin polarized solution instead, the distortion disappeared getting a perfect in plane structure, with a net magnetic moment according the g44+v.

All this is a confirmation of the exceptional features that boron offers to graphene vacancies, suppressing magnetism, saturating vacancies and di-vacancies, inducing quite different results depending of its atomic position.

In order to understand the origin of the suppression of magnetism, we proceeded comparing the band structure for different atomic configurations. This can be seen in Figure 8. In the case of g66+bfv and g66+bav_a we have two and one bands being crossed at Fermi level for up- and dn- electrons, respectively. The flat σ-band is totally occupied for up- and totally empty for dn- electrons, explaining why we observed an unequal spin filling of the bands previously mentioned.

After the structural reconstruction in g66+bav_a that leads to g66+bav_s, the flat σ-band is annihilated because of the saturation of the carbon dangling bonds, and then the partially occupied bands become equally weighted in two per spin channel as can be seen in Figure 8. Additionally, those π-bands show a remarkable dispersion, avoiding any spin polarization through Stoner's band magnetism. This is not the case for gxx+bfv and gxx+bav_a where the carbon dangling bonds are responsible for the magnetism, similar to the case of gxx+v, they add localized states to the structure and change the relative occupancies of π-bands. But still we can find differences between gxx+bfv and gxx+bav_a, since in the first case the major changes arises in a π-band repopulation with an important charge transfer from adjacent carbon atoms to boron, as was described for gxx+b. In the case of gxx+bav_a the major changes occur at the flat σ-band since boron reduces the spin-polarization of the charge density at the pentagon, position 1 and 2, because of the induced charge transfer. Even though charge transfer to boron in g66+bav_a is lower than in g66+b, 0.68 in comparison to 0.95, it is sufficient to reduce the net magnetic moment from 1.38 $\mu_B$ to 1.10 $\mu_B$ if we compare gxx+b, the situation is presented in Figure 9. In summary, boron positions and its distance to vacancy are key elements for tuning electronic properties of graphene.

## IV. CONCLUSIONS

We have discussed the role of boron doping in saturated and non-saturated graphene. The process of boron doping is unfavorable for perfect graphene, but it turns favorable when boron atoms fill carbon vacancies. As a consequence, modified graphene/graphite can be used as a reactant for doping purposes.

Regarding electronics, boron atom become negative charged when it is incorporated in the lattice, and together with its p-hole doping character for carbon, it can be used as an electronic tuning of carbon properties. We have demonstrated that magnetisms arises from the presence of dangling bonds, showing how saturation annihilates localized states and conducts to balance between the π-states, which are responsible for the extension of the magnetism to the whole sample.

In the case of magnetism our results indicate that a boron atom behaves quite different when occupies atomic position near or far from vacancies. In the first case it can migrates at the center of a single- or double-vacancy saturating bonds and the destroying the magnetism. When boron occupies a bulk position, far from a vacancy, it slightly enhances the magnetic response of the system. This feature can be used in experiments for obtaining magnetic and non-magnetic samples. Starting from modified graphite as a reactant we expected a favorable process for incorporation of boron. For high concentration of boron, we expect bulk doping and saturation of vacancies, conducting to non magnetic samples. For low concentration of boron, the amount of boron is not sufficiently to saturate all the vacancies, and now a magnetic signal could emerge from those samples, and depending on the relation of vacancies/boron there will be stronger or weaker magnetic signals.

All these findings suggest that new chemical modifications to graphene, and new type of vacancies can be used for interesting applications such as sensor and chemical labeling.

## ACKNOWLEDGMENTS

The authors thank the PEDECIBA, CSIC and Agencia Nacional de Investigación e Innovación (ANII) - Uruguayan organizations- for financial support. AWM would like to acknowledge Uruguay-INNOVA for financial support.

**Figure Caption**

Figure 1.- (a) Spin density map for g55+v indicating the sp$^2$ orbitals located at 1.0 eV below Fermi level. Look at the localization of these states in comparison with (b) the corresponding spin polarized density originated by π orbitals. The band structure for the Γ-M-K-Γ is presented in (c) indicating σ- and π- bands.

Figure 2.- Density of states showing the sp$^2$ and p$_z$ character for g22, g33, g44, g55 and g66. Finally, a graphene sketch including the nomenclature used in the text for carbon atoms near the vacancy.

Figure 3.- Mulliken's population analysis and most representative bond distances for g66+b.

Figure 4.- The diffusion path for B in the curing of a graphene vacancy, starting from the out-of-plane initial position (a) to the final in-plane disposition (d). The energy path is presented in graph (e).

Figure 5.- Band structure for: (a) g22 and g22+b including an additional electron and (b) g66 and g66+b including an additional electron. The wavefunction plotted at Dirac point for the twofold degenerated g66 graphene is presented in (c) and (d). Wavefunction plot at Dirac point for g66+b+e for zero-gap (e) and non-zero-gap (f) bands (blue and red denotes positive and negative values for the wavefunction, respectively).

Figure 6.- Optimized coordinates for: (a) graphene 5x5 with boron far from vacancy g55+bfv, (b) graphene 5x5 with boron adjacent to vacancy with asymmetric disposition g55+bav_a, (c) and (d) graphene 5x5 with boron in vacancy with symmetric disposition g55+bav_s. Carbon atoms are in brown and boron atoms in green.

Figure 7.- Vibrational density of states for g44+bav_s showing real frequencies

Figure 8.- Spin dependent band structure and corresponding density of states for: (a) g66+bfv, (b) g66+bav_a and (c) g66+bav_s. Note how the flat σ-band disappear from (a)/(b) to (c)

Figure 9.- Spin density maps for: (a) g66+bav_a and (b) g66+bfv. Blue and red colors indicate positive and negative values respectively.

**Table Caption**

Table 1.- Total magnetic moment and vacancy formation energy for different supercells with full geometrical optimization.

Table 2.- Total magnetic moment and B-doped graphene formation energy for different supercells with full geometrical optimization.

Table 3.- Total magnetic moment and formation energy for different supercells of boron doped graphene with coexistence of vacancies.

| System | $N_{atom}$ | k-points | $E_{Form}(1)$ (eV) | Magn. mom. ($\mu_B$) | Magn. mom. per C atom ($\mu_B$) |
|--------|------------|----------|--------------------|-----------------------|---------------------------------|
| g22+v  | 7          | 40x40x1  | 7.160              | 0.963                 | 0.142                           |
| g33+v  | 17         | 20x20x1  | 7.499              | 1.285                 | 0.075                           |
| g44+v  | 31         | 40x40x1  | 7.420              | 1.120                 | 0.036                           |
| g55+v  | 49         | 20x20x1  | 7.445              | 1.338                 | 0.027                           |
| g66+v  | 71         | 5x5x1    | 7.472              | 1.261                 | 0.018                           |

Table 1.-        Faccio *et al*

| System | $N_{atom}$ | $E_{Form}(2)$ (eV) | $E_{Form}(2)$ (eV) | Lattice difference (%) * | B net-charge (e) | C net-charge (e)** | Magn. mom. ($\mu B$) |
|---|---|---|---|---|---|---|---|
| g22+b | 8  | 1.193 | -5.966 | 3.8  | -0.898 | 0.600 | 0 |
| g33+b | 18 | 1.074 | -6.424 | 2.0  | -0.930 | 0.534 | 0 |
| g44+b | 32 | 0.955 | -6.463 | 1.1  | -0.934 | 0.528 | 0 |
| g55+b | 50 | 0.951 | -6.494 | 0.3  | -0.942 | 0.522 | 0 |
| g66+b | 72 | 0.919 | -6.552 | -0.1 | -0.948 | 0.522 | 0 |

* Cell parameters differences expressed as %=$(a_{gxx+b}-a_{gxx})/a_{gxx}*100$

** Total net charge for the three carbon atoms located at the first coordination shell around boron.

Table 2.  Faccio *et al*

| System | $N_{atom}$ | Magn. mom. ($\mu_B$) | $E_{norm}$ (eV) | $E_{form}(4)$ (eV) |
|---|---|---|---|---|
| g33+bfv | 18 | 1.36 | 1.207 | -0.031 |
| g33+bav_a |  | 1.16 | 0.000 | -1.238 |
| g44+bfv | 32 | 1.32 | 2.658 | 1.144 |
| g44+bav_a |  | 1.00 | 0.297 | -1.217 |
| g44+bav_s |  | 0.00 | 0.000 | -1.514 |
| g55+bfv | 50 | 1.00 | 2.599 | 1.043 |
| g55+bav_a |  | 1.00 | 0.358 | -1.198 |
| g55+bav_s |  | 0.00 | 0.000 | -1.556 |
| g66+bfv | 72 | 1.38 | 2.625 | 1.042 |
| g66+bav_a |  | 1.10 | 0.395 | -1.187 |
| g66+bav_s |  | 0.00 | 0.000 | -1.582 |

Table 3    Faccio *et al*

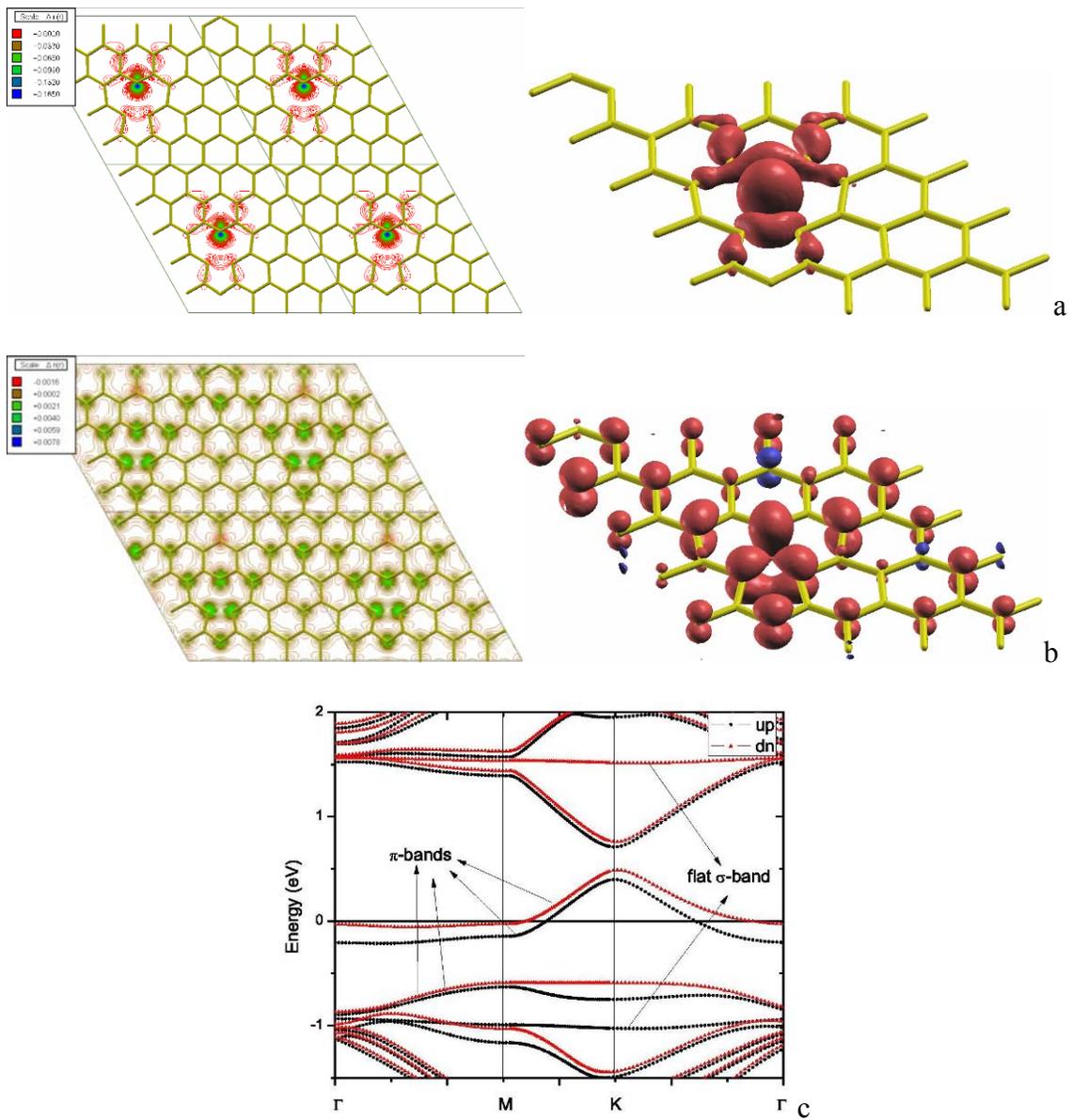

Figure 1. Faccio *et al*

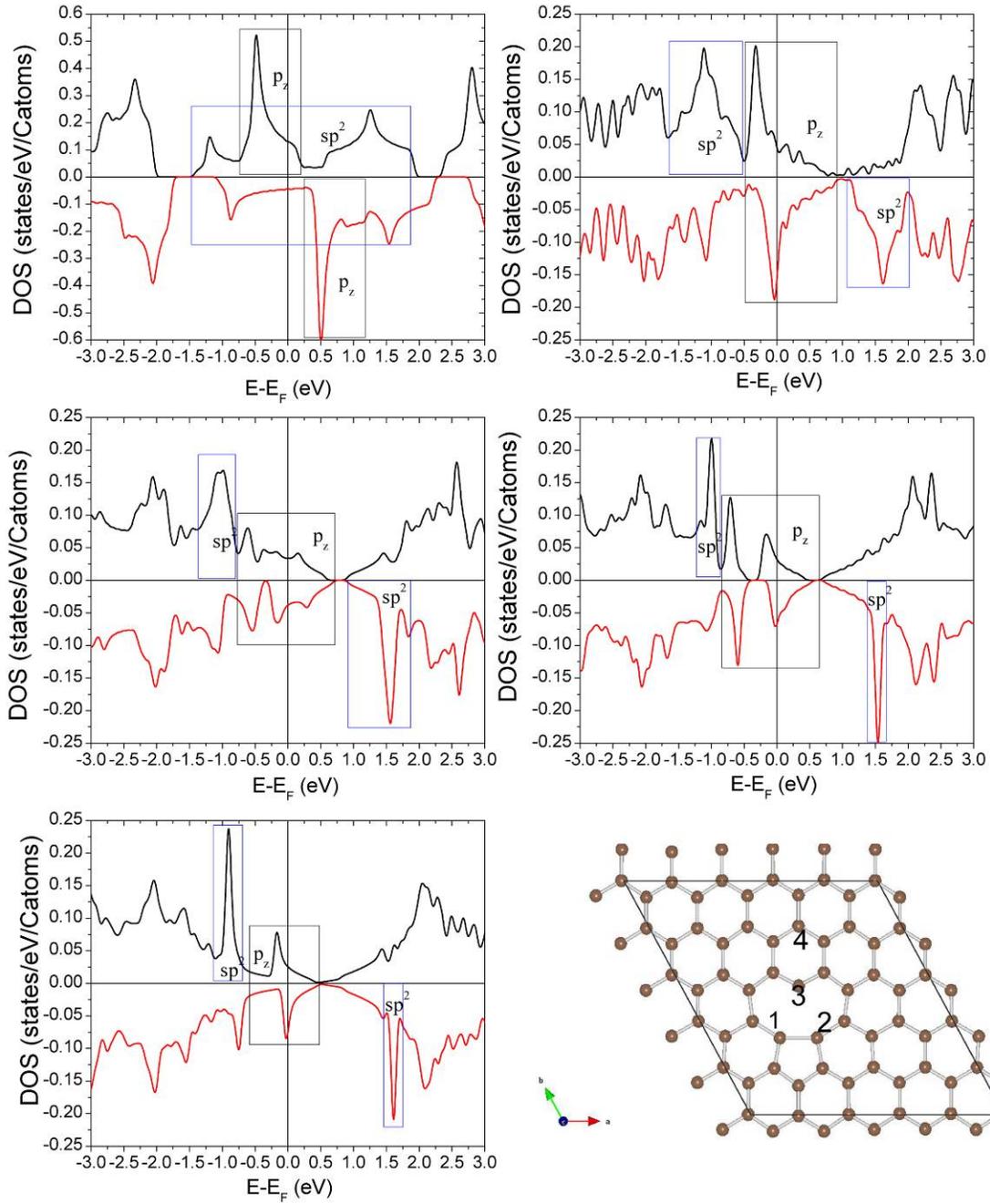

Figure 2. Faccio *et al*

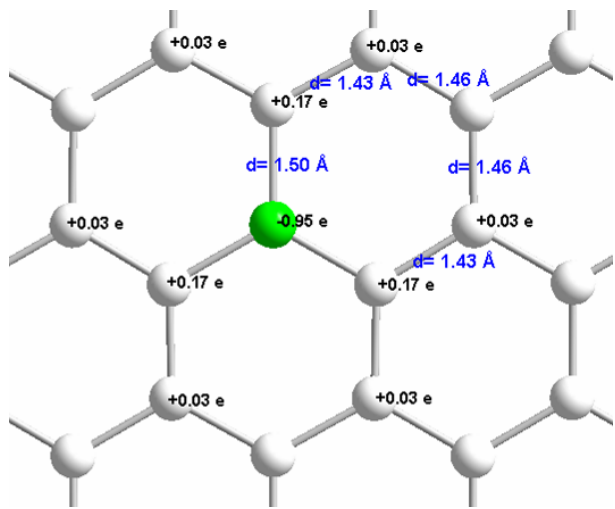

Figure 3.                          Faccio *et al*

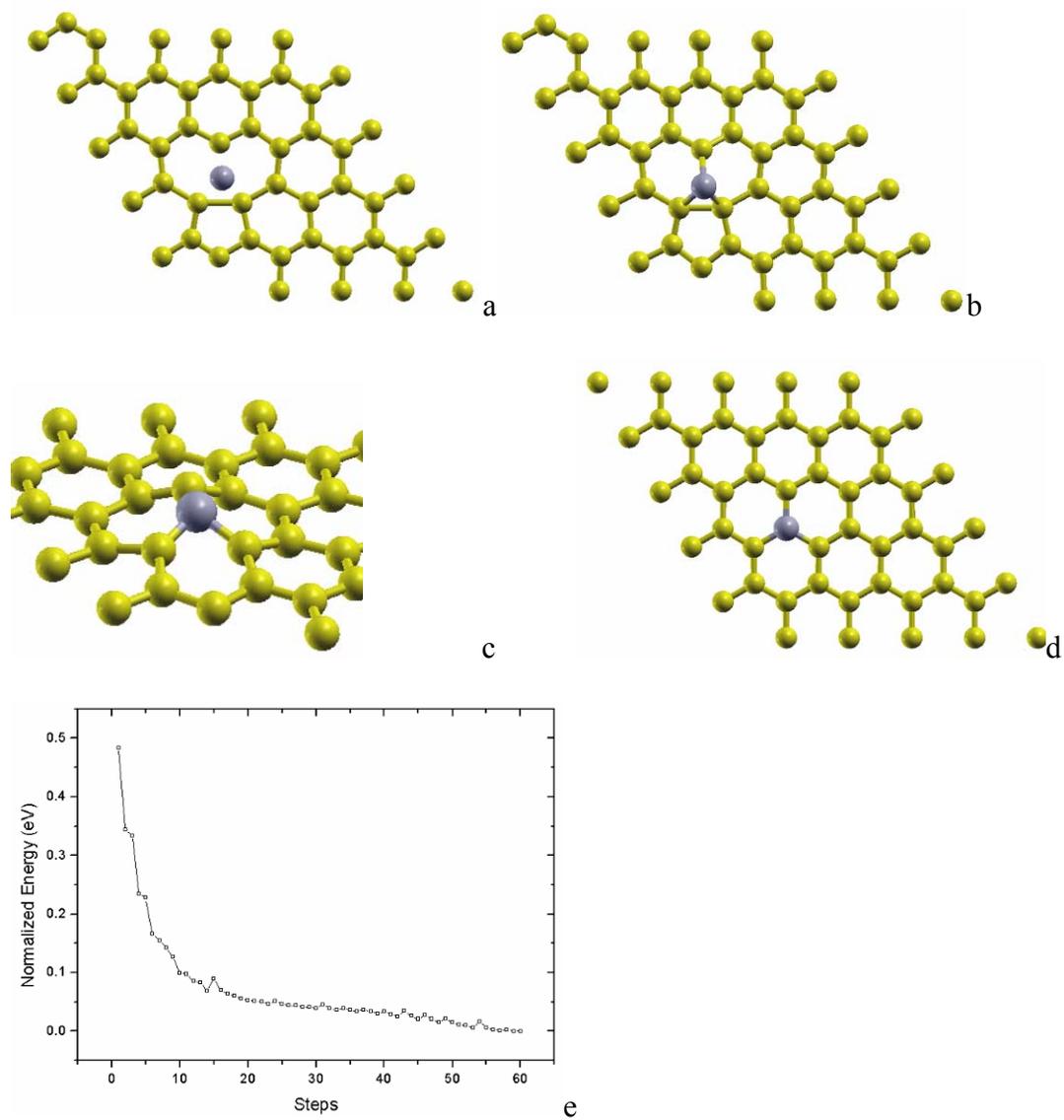

Figure 4.- Faccio *et al*

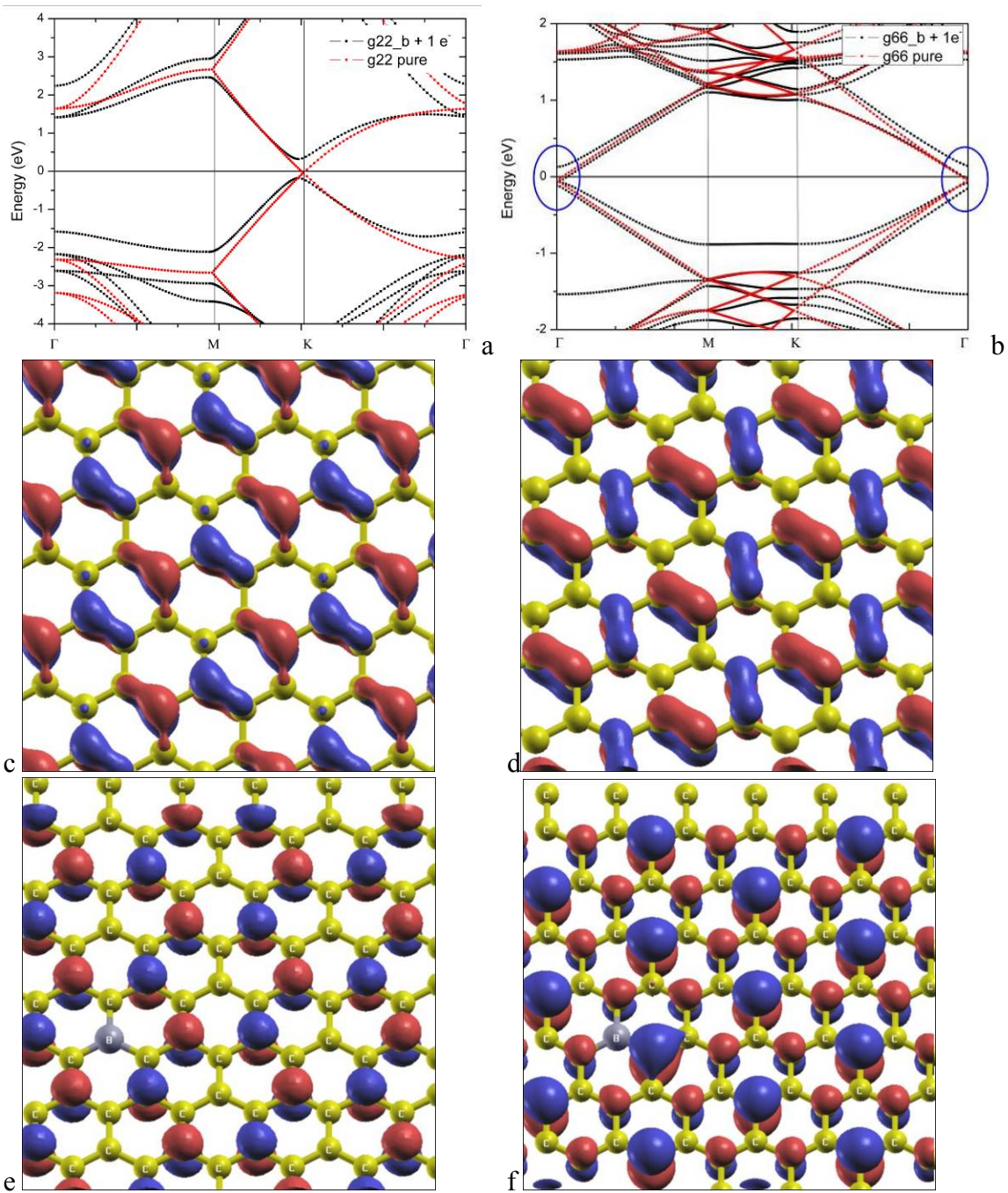

Figure 5.- Faccio *et al*

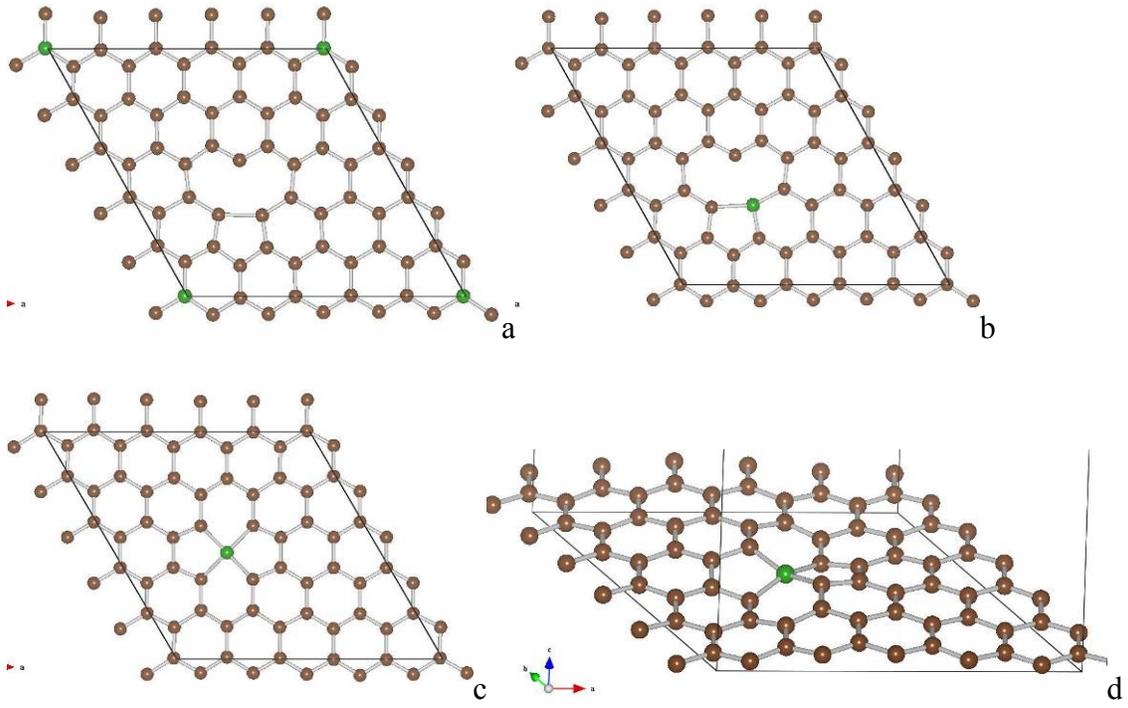

Figure 6.- Faccio *et al*

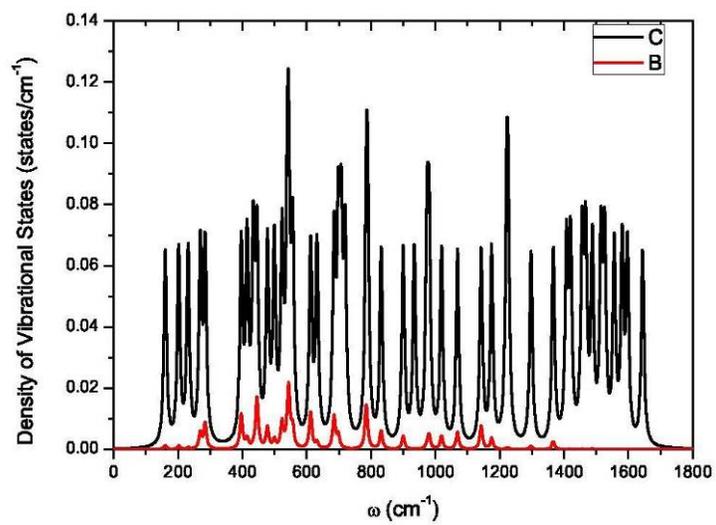

Figure 7.- Faccio *et al*

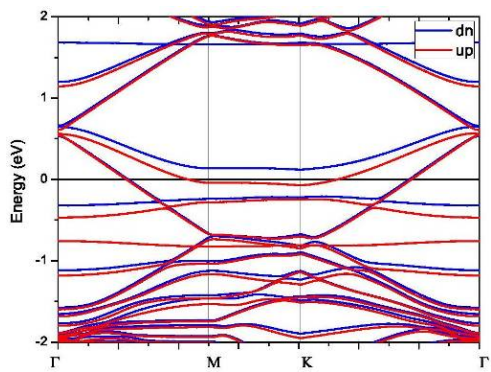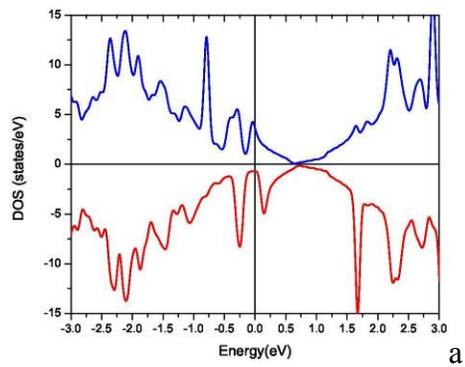

a

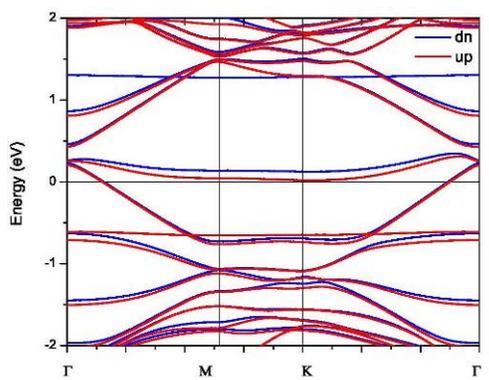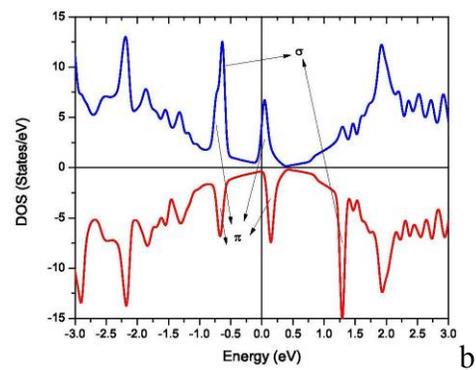

b

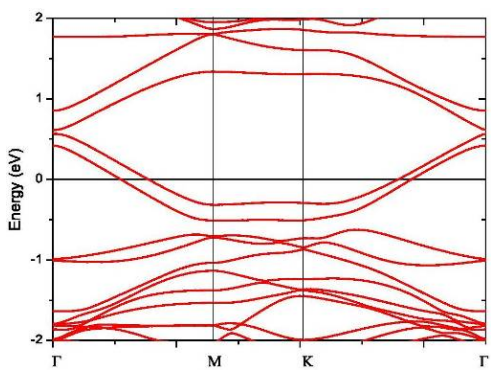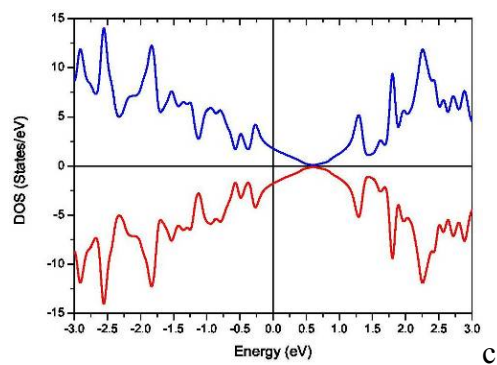

c

Figure 8. Faccio *et al*

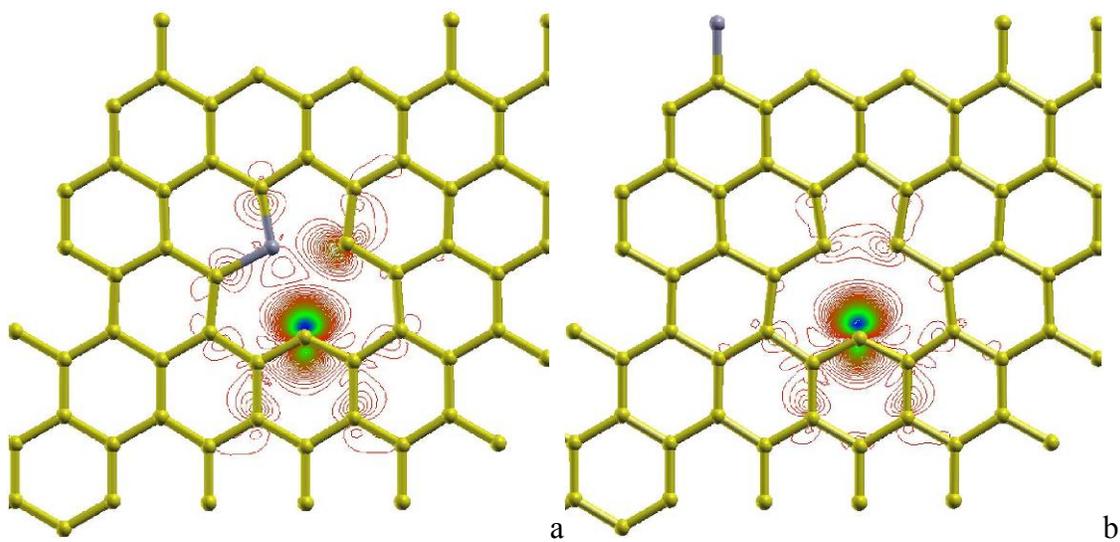

a b

Figure 9.- Faccio *et al*